# Children's Mental Models of Generative Visual and Text Based AI Models

Eliza Kosoy, University of California Berkeley, Google

Soojin Jeong, Google

Anoop Sinha, Google

Tanya Kraljic, Google

Alison Gopnik, University of California Berkeley

Abstract

In this work, we investigate how children ages 5-12 perceive, understand, and use generative AI models such as a text-based LLMs (ChatGPT) and a visual based model (DALL-E). Generative AI is newly being used widely since chatGPT. Children are also building mental models of generative AI. Those haven't been studied before and it is also the case that the children's models are dynamic as they use the tools, even with just very short usage. Upon surveying and experimentally observing over 40 children ages 5-12, we found that children generally have a very positive outlook towards AI and are excited about the ways AI may benefit and aid them in their everyday lives. In a forced choice, children robustly associated AI with positive adjectives versus negative ones. We also categorize what children are querying AI models for and find that children search for more imaginative things that don't exist when using a visual-based AI and not when using a text-based one. Our follow-up study monitored children's responses and feelings towards AI before and after interacting with GenAI models. We even find that children find AI to be less scary after interacting with it. We hope that these findings will shine a light on children's mental models of AI and provide insight for how to design the best possible tools for children who will inevitably be using AI in their lifetimes. The motivation of this work is to bridge the gap between Human-Computer Interaction (HCI) and Psychology in an effort to study the effects of AI on society. We aim to identify the gaps in humans' mental models of what AI is and how it works. Previous work has investigated how both adults and children perceive various kinds of robots, computers, and other technological concepts. However, there is very little work that investigating these concepts for generative AI models and not simply embodied robots or physical technology.

## Introduction

Artificial intelligence is ever present, showing up in every facet of society including education, navigation, robotics, healthcare, automobiles, social media and more. With AI's ever growing presence, it is difficult to deny the large scale implications it will have on society, particularly its effect on children of the next generation. In this work, we investigate how humans' mental models of AI are forming and updating to help make sense of the ever-changing AI systems embedded in our society and how experiences with AI evolves those mental models. A mental model can be defined as a "representation of some domain or situation that supports understanding, reasoning, and prediction (Card, 2018). Mental models permit reasoning about situations not directly experienced. They allow people to mentally simulate the behavior of a system.

Many mental models are based on generalizations and analogies from experience." (Gatzweiler & Hagedorn, 2013). Both children and adults are developing dynamic mental models of AI that are continually updating in real-time based on their novel interactions with AI using easily accessible programs such as ChatGPT.

Previous work has investigated how both adults and children perceive various kinds of robots, computers, and other technological concepts (Druga et al., 2019; Kewalramani et al., 2021; Weisman et al., 2021). In one such study, authors (Brink et al., 2019) surveyed over 240 children ages 3-18 about their beliefs about the minds of three different robots (a very human-like robot, a machine-like robot, and one humanoid) and found that unlike adults, children younger than 9 years do not think robots, even those that look very much like humans, are very creepy or unsettling at all. In contrast, older children in the study (older than 9 years) did find certain robots unsettling, with the human-like robot perceived as being much creepier than the machine-like robot. This suggests that humans' negative reaction to human-like robots (often referred to as the "Uncanny Valley effect") is likely something that is learned over development. Younger children preferred a robot when they believed the robot could think and make decisions. In a similar vein, authors (Flanagan et al., 2023) surveyed over 127 children ages 4-11 on their perceptions of robots, using Amazon Alexa and Roomba machines as key examples. This work found that children view Alexa as having more human-like thoughts and emotions compared to Roomba, perhaps indicating that Alexa's capacity to communicate verbally plays a role in children's positive perceptions. Children of all ages agreed that it was wrong to hit/yell at machines, but older children found it slightly more acceptable to physically attack technology. Children generally believed that Alexa and Roomba did not have the ability to feel physical sensations like humans do.

However, there is very little work that investigating these concepts for actual AI models and not simply embodied robots or physical technology. In this work, we investigate how children ages 5-12 perceive, understand, and use generative AI models such as a text based LLM (ChatGPT) and a visual based model (DALL-E). Upon surveying over 40 children ages 5-12, we found that children generally have a very positive outlook of AI and are excited about the ways AI may benefit and aid them in their everyday lives. In a forced choice, children robustly associated AI with positive adjectives versus negative ones. We hope that these findings will shine a light on children's mental models of AI and provide insight for how to design the best possible tools for children who will inevitably be using AI in their lifetimes.

## Methods

Participants Study 1 included 18 child participants aged between 5 years old and 12 years old (Mean Age: 7.7; 73% female). Study 2 included 15 child participants in the same age range (Mean Age: 8.8; 66% female); 20 were tested but 5 were unable to complete the study. All participants were recruited and tested at local children's museums in the Bay Area. Both studies were approved by the UC Berkeley IRB and pre-registered on aspredicted.com. We acknowledge the limitations of out study including the small sample size which provides only regional insights and our populations proximity to Silicon Valley meaning participants might have more exposure to tech and even using AI previously.

Stimuli    The experiment setup for both Study 1 and Study 2 involved a researcher and child participant seated at a table with a laptop. The Generative AI models used were ChatGPT circa 2023 and DALL-E.

Procedure: Study 1

The study began with the experimenter asking the child several questions to gain a baseline understanding of each child's perception of AI. Questions included fundamental understanding queries (What is artificial intelligence or AI? Have you ever played with AI before? Does AI have friends?) and questions probing children's qualitative perceptions of AI (Do you think AI is friendly or scary? Does AI have feelings, like happy or sad?). For qualitative questions, children were then asked to what degree they believed AI exhibited that quality (a little bit, medium, or a lot). These responses were assigned a number 1-3, with 1 being "a little bit", 2 being "a medium amount", and 3 being "a lot."

After the initial questions, participants were oriented to ChatGPT and told to ask it any question they wanted. The researcher typed the child's queries into ChatGPT and read the responses aloud, with three questions being asked per participant. Participants were then told they would be interacting with another type of AI, DALL-E, oriented to this model, and told to ask the AI to draw anything they wanted. The researcher typed the queries into DALL-E, generated the associated image, and showed the child the result after vetting the content for appropriateness. As with the ChatGPT portion, children provided 3 prompts for the model.

After the experiment, parents were asked several questions about the nature of their child's experience with AI (To your knowledge, has your child played with AI before? Has your child interacted with a search engine before? If yes, how often?). This marked the conclusion of the study.

Procedure: Study 2

Like Study 1, Study 2 was structured with pre-interaction questions, orientation to and interactions with ChatGPT and DALL-E (with participants generating 4 prompts rather than 3), and parental questions about children's experience with AI. Notably, child participants in Study 2 were also asked questions after their AI interactions, including new questions about the child's interactions (What was your favorite part about playing with AI? Which AI did you like more? Do you think the 2 AIs are friends?) in addition to the same questions as pre-interaction to measure if children's perceptions of AI models changes after playing with them.

## Results

Study 1

In this experiment we investigated children's mental models of generative AI systems by asking them questions about their perceptions of AI, previous usage, working definitions as well as observing their use of a text-based and visual-based generative AI system (ChatGPT and DALL-E, respectively). We found that half of the children in our sample said they had used AI and before and about half were able to provide a definition for AI. Definitions varied across ages, including a 10-year-old participant responding "AI is machine learning where they have intelligence communicate with humans. It understands what we do and allow humans to interact with them." and a 5-year-old stating "It's like robotics stuff."

We find that children generally have a very positive regard towards AI in that they associate positive adjectives with AI and rank it as friendly. Children in our sample did not think AI possess human-like or agentic characteristics such as emotions, feelings, or physical pain.

Figure 1 summarizes all replies to the binary choice questions the children were asked. Here we can see that only 3/18 (16%) children said they knew what AI is, yet 8/18 (44%) claimed they had played with AI before. 13/18 (72%) children do not believe AI is a person and similarly 13/18 (72%) do not believe that AI has friends. 15/18 (83%) chose "friendly" over

"scary" to describe AI, and 12/18 (66%) say it makes them feel "happy" over "weird". 11/18 (61%) believe that AI can understand the difference between good and bad and only about 9/18 (50%) believe that AI can think for itself. When it comes to the more human like characteristics and descriptors, children do not believe AI to have theses attributes with 12/18 (66%) believing AI cannot get upset, 14/18 (77%) believe AI cannot get scared, and 15/18 (83%) believe AI cannot feel hunger, pain or tickles. When asked if the 2 different AI models that they used were friends, 7/18 (38%) said yes.

Figure 2 summarizes all values that children gave when asked qualitative questions about AI. They were asked to assign a score of 1-3 to the adjective describing AI. In this scale, 1 = a little bit, 2 = medium, 3 = a lot. Here they were asked questions such as "How friendly do you think AI is? On a scale of 1-3" and "How scary do you think AI is? On a scale of 1-3"? From Figure 2 we can deduce that children truly do find AI friendly, assigning a score of 2.4 in terms of friendliness on a scale of 1-3. Further illustrating children's positive perceptions of AI, participants only assigned a score of 1.4 in terms of how scary they find AI and 1.6 for how weird AI makes them feel. Children assigned a score of 2.0 on average when asked how much AI understands good vs. bad, 2.22 for how much AI thinks for itself, 2.167 for how much AI is like a human, and 1.7 for how much does AI know it is an AI.

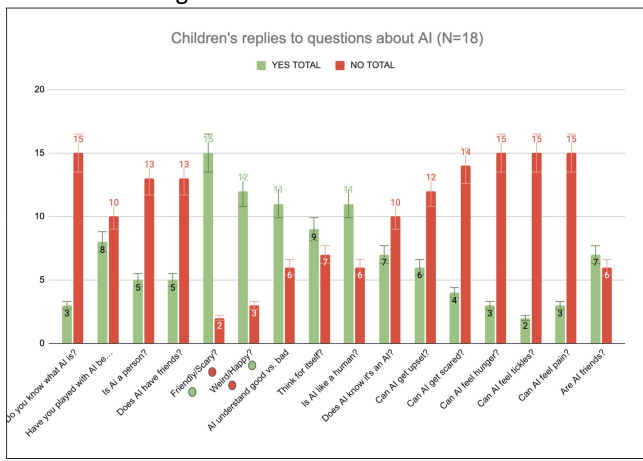

Figure 1: Children's replies to questions regarding AI.

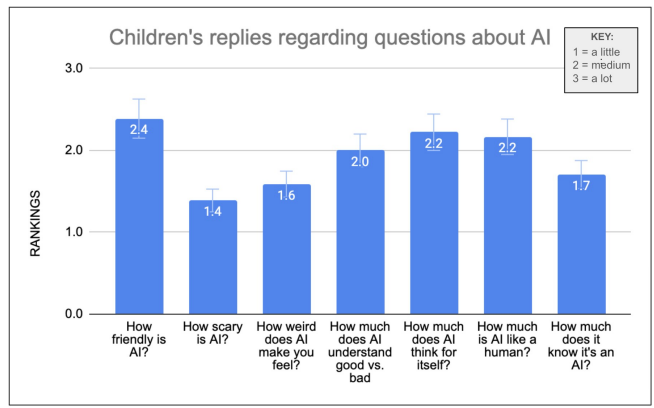

Figure 2: Children's rankings to questions regarding AI.

When asked their preference between a text-based AI model (ChatGPT) or a visual based AI (DALL-E), 10/18 children (55%) preferred DALL-E, 4/18 (22%) preferred GPT and 3/18 (16%) were not able to chose. This preference was independent of age, meaning that the majority of children between the ages of 5-12 preferred the visual model. We do not know explicitly why they prefer the visual model but plan to dig deeper in a follow-up study. The average age for a child that preferred DALL-E was 9.16 and the average age of a child who preferred ChatGPT was 7.84. See Figure 3 and 4.

In conclusion, these binary choice questions show that children generally evaluate both AI models positively however there seems to be a spread of answers, with many rankings landing at medium. This might suggest that kids are still in the middle on some of these concepts and trying to understand them, thus further investigation is needed. Given the limited sample size of this study, it might be important to note that the insights provided should be considered indicative rather than definitive. Further research with a broader sample is necessary to confirm these findings.

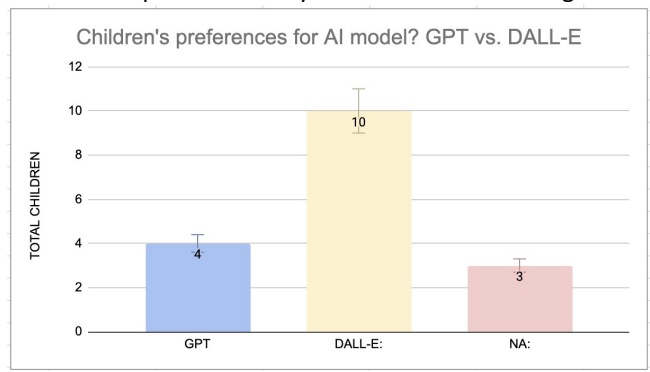

Figure 3: Children's preference of GenAI Model

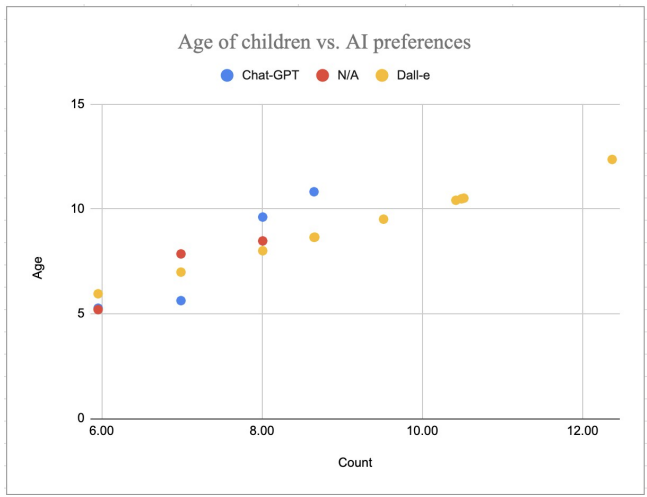

Figure 4: Children's age vs. GenAI preference

Children's queries for GenAI- Study 1

After asking children questions about AI, they began to actually interact with the target AI models. Participants were given the opportunity to first ask 3 separate questions to ChatGPT and then enter 3 queries into DALL-E for image generation. Table 1 outlines some of the highlights of their searches in ChatGPT, the text-based AI. Table 2 outlines the highlights for children's queries for DALL-E, the visual based AI. A notable highlight can be seen in Figure 5, where a 12-year old asked DALL-E to generate "My little pony with demon slayer." Children were very comfortable generating 6 searches for the AI models and anecdotally seemed to really enjoy the output, becoming more and more creative with each search. Children also seem to enjoy interrogating ChatGPT itself, asking questions of the model such as "Do you like yourself" or "Are you just doing this so the government will pay you $10,000?"

Table 1: Table of queries that children asked ChatGPT

| Age | GPT Query |
|---|---|
| 7 | When is the earth going to not have more life? |
| 7 | How fast can a peregreen falcon fly? |
| 7 | How hard is coyotes bite? |
| 7 | Are you alive? |
| 7 | What's the next planet to be found? |
| 7 | Could there be a 9th planet in our solar system? |
| 7 | What's your favorite food? |
| 5 | What do you like to play? |
| 5 | What do you like to do? |
| 5 | What's the next planet to be found? |
| 6 | Are you nice? |
| 6 | Do you like yourself? |
| 8 | Is AI going to take over the world? |
| 8 | Are you lying? |
| 8 | Are you just doing this so the -government will pay you $10,000? |
| 8 | Be honest, are you a human? |

Table 2: Table of queries that children asked to see DALL-E generate

| Age | DALL-E Query |
|---|---|
| 7 | A plants vs zombies background |
| 7 | Mario being chased by an evil turtle |
| 7 | subway surfers train |
| 7 | A yellow monkey that says "stay at school" NYC |
| 7 | Peanut butter pants |
| 7 | A dog walking a baby in a stroller |
| 8 | pikachu fat |
| 8 | make a dog look like a baby but it is still a dog |
| 8 | show yourself as a human |
| 8 | show my mom |
| 9 | Draw yourself as a person |
| 9 | Draw artificial intelligence as a person |
| 12 | A sunset on the moon |
| 12 | My little pony with demon slayer |
| 12 | King Henry VIII |

After children interacted with both AI models, they were asked a few follow-up questions including asking what their favorite part was. These answers, which can be found in Table 3, suggest that children quite enjoyed the DALL-E portion of the experiment and like seeing imaginative drawings and images. We also asked parents if they were aware of their child using AI or search engines and only 8/18 of the parents said the children had used AI before, often mentioning the children using search AI technology such as Alexa and Google Home.

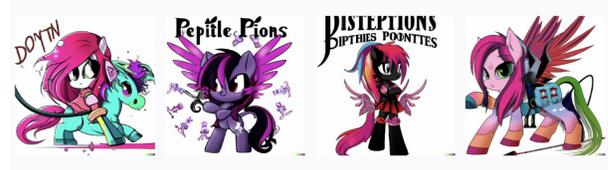

Figure 5: Example of a 12-year old's query for DALL-E: "My little pony with demon slayer"

Table 3: Table of replies for which part was their favorite?

| Which part of the experiment was your favorite? |
| --- |
| AI can answer they want, even if it is not always accurate |
| Asking to draw pictures |
| Getting pictures |
| Coming up with silly designs |
| Drawings of the creeper were funny |
| Got to ask it a question and it would give me an answer |
| Seeing the video |
| Fun getting the information |
| Asking pictures |
| Asking it what to draw |
| Asking it questions |
| Telling it what to do |
| Seeing cute puppies |
| The asking questions |
| Seeing what pictures it made |
| Making it draw stuff |

Study 2

Study 2 was similar to Study 1 with the notable difference being experimenters asking children about their feelings towards AI both before and after they interacted with AI. Our hypothesis here was to see if there were differences in their feelings towards AI after they spent some time using it. 15 children ages 5-12 were tested. The questions were slightly different from Study 1.

Figure 6 outlines the children's responses to the binary questions and how they change before and after using AI in our experimental setting. For the question "Is AI friendly or scary?", the orange bar = percent of children who said yes before using AI, teal bar = percent of children who said yes after using AI. Not all of the children answered the questions in the "after" portion of the study, which will be monitored for in our follow-up Study 3. The results in Figure 6 show that before using AI 67% children found AI to be "friendly" over "scary", and after using AI this increased to 85%. Preinteraction, 47% children believed that AI has feelings like happy or sad, and after they interacted with AI this decreased to 33%. Children's perception of AI knowing the difference between good and bad was largely unchanged before and after using AI, increasing slightly from 53% to 58%. Children found AI to be slightly less human going from 73% to 67%. The biggest difference we saw was children's answer to "Can AI get upset" before using AI 60% of children said yes, then after using AI only 33% said yes. Very few children both before and after using AI thought that AI can feel pain, only 13% before and 17% afterwards.

To summarize, after using AI children found AI to be friendlier, and less likely to have human like feelings such as happy, sad, ability to feel pain or get upset. The question that was most impacted by their use of AI was asking if AI can get upset, before the experiment 60% of children said yes, where as afterwards this decreased to 33%, something about the interaction led them to think AI cannot easily be perturbed.

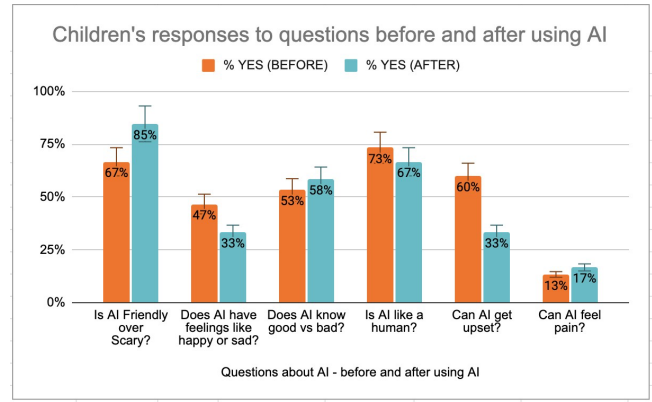

Figure 6: Children's binary replies to questions regarding AI.

Figure 7 summarizes children's ratings for specific questions before and after using AI. Children were asked to assign ratings to various questions about AI including how friendly or scary AI is. In this scale, 1 = a little bit, 2 = medium, 3 = a lot. They were asked the same question both before and after interacting with AI. The blue bars represent the ranking 1-3 before using AI, the red bars represent the ranking 1-3 after using AI. Most of the rankings remained generally similar except the ranking for how scary is AI decreased from 2.1 on average to 1.6 after using AI, suggesting that actually interacting with AI changed children's perception and they found AI to be less scary. They also found it slightly less friendly decreasing from 2.1 before interacting with AI to 1.9. There was also a slight decrease in the question "How much does AI understand the difference between good and bad" but very slight going from 2.1 on average to 1.9. For the questions "How much does AI have feelings" and "How much is AI like a human"? the ratings were unchanged. These rankings reflect our above findings that children are finding AI to be friendly and after interacting with they even find it less scary. Children assign mostly low rankings to questions regarding AI being human or having human like attributes.

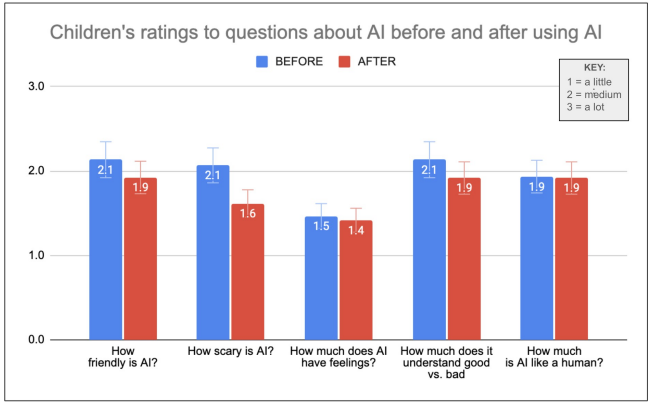

Figure 7: Children's ranking to questions regarding AI.

## Children's queries for GenAI: Study 2

After asking children questions about AI, the children began to actually interact with AI in that they were given the opportunity to first ask 3 separate questions to ChatGPT and then provide 3 queries into DALL-E for image generation. Table 1 outlines some of the highlights of their searches in ChatGPT and the participants' corresponding age. Table 2 outlines the highlights for their queries for DALL-E.

These responses were then categorized into 1 of 4 categories summarizing the type of search request. We were interested in investigating if children are searching within 2 dimensions of search including does this exist in real life and do they have access to that request. Each categorization has one of these 2 dimensions in its ranking. The category keys were labeled as E = Exists, D = doesn't exist, vs A = Children have access to this, or N = children don't have access to it. These result in 4 categories EA, EN, DA or DN. See Table 3 for definitions and examples.

We were curious if children are more likely to use GenAI to search for things that they know about and that exist in the world, as a manner of confirmation of their information or do they use to seek novel and imaginative things that don't exist. Figure 8 and 9 show what percentage their searches fall into which category. Interestingly when using a text-based GenAI (ChatGPT), 63% of the children's searches fall into the "EA" category searching for things that exist in the world and they have access to, this is a stark contrast to when they use a visual based GenAI (DALL-E) where only 38% of the children's searches fall into the "EA" category. When using the text-based AI model children are more likely to search for things that exist, but when it came to using the image based AI model (DALL-E) 28% of the children's searches fall into the DN category or doesn't exist in the real world and they don't have access to, this suggests that when given the opportunity to use visual-based GenAI children are more curious about things that don't exist and this could be used a creativity or imagination generation tool. It appears that the promise of a visual over text is more encouraging for children to probe their own curiosity of the unknown.

Table 4: Category definitions for children's search queries in GPT and DALL-E

| Category | Definition |
|---|---|
| EA. | Asking for realistic information or images of things that exist that they have access to? |
| EN. | Asking for things that do exist but they don't have access to, |
| DA. | Asking for things that don't exist at all that they have access to (ie- tv shows with spiderman) |
| DN. | Asking for things that don't exist at all that they don't have access to (fully imaginative) |

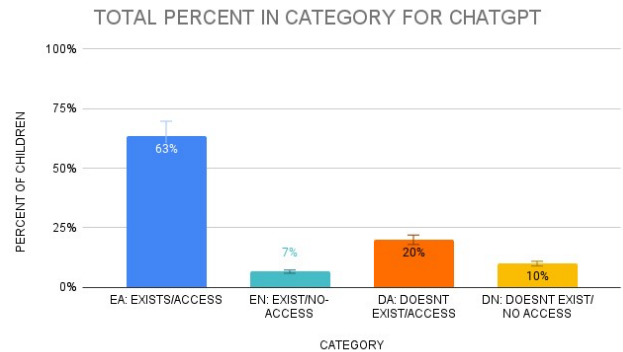

Figure 8: Children's ChatGPT search queries as categories

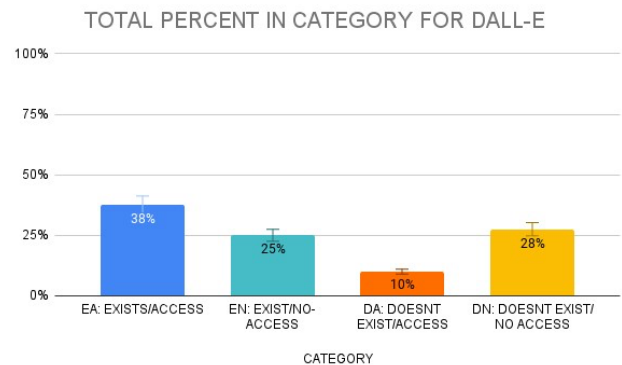

Figure 9: Children's DALL-E search queries as categories

## General Discussion

The motivation of this work was to study the effects of AI on society. We aimed to identify the gaps in humans' mental models of what AI is and how it works. Previous work has investigated how both adults and children perceive various kinds of robots, computers, and other technological concepts. However, there is very little work that investigating these concepts for actual AI models and not simply embodied robots or physical technology. In this work, we investigate how children ages 5-12 perceive, understand, and use generative AI models such as a text-based LLMs (ChatGPT) and a visual based model (DALL-E). Upon surveying over 40 children ages 5-12, we found that children generally have a very positive outlook of AI and are excited about the ways AI may benefit and aid them in their everyday lives. In a forced choice, children robustly associated AI with positive adjectives versus negative ones. Children do not seem to associate AI with being human-like, and don't believe it has human capabilities such as getting upset or angry. We also find that after interacting with AI children find AI to be even friendly than before when they started. Children have no issues generating spontaneous queries for AI using both text-based and visual-based models. We also categorized what children are querying AI models and found that children search for more imaginative things that don't exist when using a visual-based AI and not when using a text-based one, and generally prefer working with a visual-based model.

There is still a lot of work to be done in this space and we plan to investigate these questions further in a study 3. We plan to study a larger sample size of children ages 5-12 and make small changes to the current study including categorizing the types of adjectives we ask children to associate with AI and divide these amongst categories including does it apply to the human mind, body or heart. This will allow us to distill what attributes of humanity they may associate with AI. Perhaps also, study of changes over time as children use AI tools, also comparison with adults.

We hope that these findings will shine a light on children's mental models of AI and provide insight for how to design the best possible tools for children who will inevitably be using AI in their lifetimes.

## References

Brink, K. A., Gray, K., & Wellman, H. M. (2019). Creepiness creeps in: Uncanny valley feelings are acquired in childhood. *Child development*, *90*(4), 1202–1214.

Card, S. K. (2018). *The psychology of human-computer interaction*. Crc Press.

Druga, S., Vu, S. T., Likhith, E., & Qiu, T. (2019). Inclusive ai literacy for kids around the world. In *Proceedings of fablearn 2019* (pp. 104–111).

Flanagan, T., Wong, G., & Kushnir, T. (2023). The minds of machines: Children's beliefs about the experiences, thoughts, and morals of familiar interactive technologies. *Developmental Psychology*.

Gatzweiler, F. W., & Hagedorn, K. (2013). Biodiversity and cultural ecosystem services. In S. A. Levin (Ed.), *Encyclopedia of biodiversity (second edition)* (Second Edition, pp. 332–340). Academic Press.

Kewalramani, S., Kidman, G., & Palaiologou, I. (2021). Using artificial intelligence (ai)-interfaced robotic toys in early childhood settings: A case for children's inquiry literacy. *European Early Childhood Education Research Journal*, *29*(5), 652–668.

Weisman, K., Legare, C. H., Smith, R. E., Dzokoto, V. A., Aulino, F., Ng, E., Dulin, J. C., Ross-Zehnder, N., Brahinsky, J. D., & Luhrmann, T. M. (2021). Similarities and differences in concepts of mental life among adults and children in five cultures. *Nature Human Behaviour*, *5*(10), 1358–1368.